\let\ifarxiv=\iftrue     % ARXIV VERSION
\pdfoutput=1

\ifarxiv

\documentclass[12pt,a4paper]{article}
\usepackage[a4paper,text={450pt,650pt},centering]{geometry}

\fi

\ifarxiv\else

\documentclass[11pt,a4paper]{article}
\usepackage{mathptmx}
\usepackage[a4paper,text={130mm,198mm}]{geometry}

\fi

\usepackage{amsmath,amssymb}
\usepackage[bookmarks=true,hyperfigures=true]{hyperref}
\usepackage{graphicx}
\usepackage[nosort]{cite}
\ifarxiv\usepackage[bulletsep]{collref}\fi
\let\oldbfseries=\bfseries
\let\oldmdseries=\mdseries
\let\oldnormalfont=\normalfont
\renewcommand{\bfseries}{\oldbfseries\boldmath}
\renewcommand{\mdseries}{\oldmdseries\unboldmath}
\renewcommand{\normalfont}{\oldnormalfont\unboldmath}

\allowdisplaybreaks[3]

\numberwithin{equation}{section}

\usepackage[font=small,labelfont=bf,width=0.85\textwidth]{caption}

\providecommand{\hypersetup}[1]{}
\providecommand{\texorpdfstring}[2]{#1}

\hypersetup{plainpages=false}
\hypersetup{pdfpagemode=UseNone}
\hypersetup{bookmarksnumbered=true}
\hypersetup{pdfstartview=FitH}
\hypersetup{colorlinks=false}
\hypersetup{citebordercolor={.5 1 .5}}
\hypersetup{urlbordercolor={.5 1 1}}
\hypersetup{linkbordercolor={1 .7 .7}}
\providecommand{\href}[2]{#2}
\providecommand{\arxivlink}[1]{\href{http://arxiv.org/abs/#1}{arxiv:#1}}

\newcommand{\h}{\hat}
\renewcommand{\t}{\tilde}
\def\Gslash{ \, {\relax{/ \kern-.54em  G}}}
\def\slashint{\relax{\rm \hbox{-}\hbox{-} \kern-1.05em \int}}
\def\Slashint{\relax{\rm - \kern-.9em \int}}
\def\pslash{ \, {\relax{/ \kern-.55em  p}}}
\def\qslash{ \, {\relax{/ \kern-.55em  q}}}

\begin{document}

%%%%%%%%%%%%%%%%%%%%%%%%%%%%%%%%%%%%%%%%%%%%%%%%%%%%%%%%%%%%%%%%%%%%%%%%%%%%%%%%
%%%%%%%%%%%%%%%%%%%%%%%%%%%%%%%%%%%%%%%%%%%%%%%%%%%%%%%%%%%%%%%%%%%%%%%%%%%%%%%%
% TITLE PAGE

\thispagestyle{empty}
\phantomsection
\addcontentsline{toc}{section}{Title}

\begin{flushright}\footnotesize%
\texttt{kcl-mth-10-18},
\texttt{\arxivlink{1012.3989}}\\
overview article: \texttt{\arxivlink{1012.3982}}%
\vspace{1em}%
\end{flushright}

\begingroup\parindent0pt
\begingroup\bfseries\ifarxiv\Large\else\LARGE\fi
\hypersetup{pdftitle={Review of AdS/CFT Integrability, Chapter II.4: The Spectral Curve}}%
Review of AdS/CFT Integrability, Chapter II.4:\\
The Spectral Curve 
\par\endgroup
\vspace{1.5em}
\begingroup\ifarxiv\scshape\else\large\fi%
\hypersetup{pdfauthor={Sakura Schafer-Nameki}}%
Sakura Sch\"afer-Nameki
\par\endgroup
\vspace{1em}
\begingroup\itshape
Kavli Institute for Theoretical Physics, University of California, \\
Santa Barbara, CA 93106, USA\\ 
and  \\
Department of Mathematics, King's College, \\
The Strand, London, WC2R 2LS, UK
\par\endgroup
\vspace{1em}
\begingroup\ttfamily
ss299 theory.caltech.edu
\par\endgroup
\vspace{1.0em}
\endgroup

\begin{center}
\includegraphics[width=5cm]{TitleII4.mps}%figure for your chapter
\vspace{1.0em}
\end{center}

\paragraph{Abstract:}
We review the spectral  curve for the classical string in $AdS_5 \times S^5$. Classical integrability of the $AdS_5\times S^5$ string implies the existence of a flat connection, whose monodromies generate an infinite set of conserved charges. The spectral curve is constructed out of the quasi-momenta, which are eigenvalues of the monodromy matrix, and each finite-gap classical solution can be characterized in terms of such a curve. This provides a concise and powerful description of the classical solution space. In addition, semi-classical quantization of the string can be performed in terms of the quasi-momenta. We review the general frame-work of the semi-classical quantization in this context and exemplify it with the circular string solution which is supported on $ \mathbb{R}  \times S^3\subset AdS_5 \times S^5$.

\ifarxiv\else
% http://www.ams.org/msc
\fi
\hypersetup{pdfsubject={MSC (2010): 81T30, 81T60, 70H06, 14H70}}%

\ifarxiv\else
\fi
\hypersetup{pdfkeywords={AdS/CFT, Integrability, Spectral Curve, Semi-classical string quantization}}%

\newpage

%%%%%%%%%%%%%%%%%%%%%%%%%%%%%%%%%%%%%%%%%%%%%%%%%%%%%%%%%%%%%%%%%%%%%%%%%%%%%%%%
%%%%%%%%%%%%%%%%%%%%%%%%%%%%%%%%%%%%%%%%%%%%%%%%%%%%%%%%%%%%%%%%%%%%%%%%%%%%%%%%
% BODY

%%%%%%%%%%%%%%%%%%%%%%%%%%%%%%%%%%%%%%%%%
%%%%%%%%%%%%%%%%%%%%%%%%%%%%%%%%%%%%%%%%%
\tableofcontents
\section{Introduction and Outlook}

The integrability of the classical superstring in $AdS_5 \times S^5$ follows from the existence of an infinite set of conserved charges \cite{Bena:2003wd}. In principle this allows for a complete classical solution of the theory, albeit in practice finding explicit classical solutions may be limited to simple field configurations. However, in the context of the spectral AdS/CFT correspondence, where the main objective is to map the spectrum of string energies in $AdS_5 \times S^5$ to the spectrum of anomalous dimensions of four-dimensional $\mathcal{N}=4$ Super-Yang Mills (SYM), finding explicit solutions is not of primary interest. Indeed, it is much more important to find a way to directly characterize the spectrum. On the SYM theory side, this was achieved by noting that certain Bethe ans\"atze compute the spectrum of the dilatation operator. For the dual classical and semi-classical string theory in $AdS_5\times S^5$ this role is played by the spectral curve.
 
More specifically, using the classical Lax connection \cite{Bena:2003wd} and the monodromy matrix obtained by parallel transporting the connection around the worldsheet,  it is possible to setup an elegant framework, which allows to characterize all finite gap solutions in terms of complex algebraic curves.
In this geometric description, finite-gap translates into finite genus of the curve. The conserved charges, such as  the energy, can in this way be computed without having to solve the equations of motion. 
Furthermore, semi-classical quantization can be described in this framework, and allows for a concise description of the one-loop energy shifts presented in the part of the review \cite{chapQString}.

The seminal paper \cite{Kazakov:2004qf} was the first to point out the importance of the spectral curves for the integrable systems that arise in the AdS/CFT correspondence. The algebraic curves for the classical string in the $\mathbb{R}\times S^3$ subspace and the corresponding subsector of one-loop planar $\mathcal{N}=4$ SYM were shown to agree by some simple identifications. On the gauge theory side, the spectral curve emerges in the thermodynamic limit  of the  ferromagnetic Heisenberg-spin chain, that diagonalizes the one-loop dilatation operator in the $\mathfrak{su}(2)$ subsector.  Subsequently, this analysis was generalized to the $\mathfrak{sl}(2)$ subsector or $AdS_3 \times S^1$ string solutions \cite{Kazakov:2004nh}, the $\mathfrak{su(4)}$ subsector \cite{Beisert:2004ag} and finally to the complete $\mathfrak{psu} (2,2|4)$ symmetric one-loop Heisenberg spin-chain \cite{SchaferNameki:2004ik, Beisert:2005di} and the $AdS_5 \times S^5$ superstring \cite{Beisert:2005bm}. 

Apart from providing a nice geometric description of classical solutions to the superstring, or in the dual theory, Bethe root configurations in the thermodynamic limit, the spectral curve is a very powerful tool to compute quantum corrections to classical string solutions. This was first advocated in the papers \cite{Beisert:2003ea, Beisert:2003xu, Beisert:2005bv} and then applied to the classical string spectral curve in \cite{Gromov:2007aq, Gromov:2007cd, Gromov:2007ky, 
Gromov:2008ec, Gromov:2008ie, Gromov:2008en, Vicedo:2008jy, Gromov:2009zz}, in particular allowing a test of the asymptotics Bethe ansatz \cite{Beisert:2006ez, Beisert:2006ib} and an explicit formula for the one-loop energy shift for a large class of solutions. 

There are various interesting questions where spectral curves should either be useful or give a more elegant description, in the context of the AdS/CFT correspondence. 
Albeit, applications to higher order $\alpha'$ corrections seem to be difficult to describe. Both conceptually and computationally, it would be very important to find a suitable all-loop quantization of the algebraic curve. To an extent, the Bethe ansatz, and more recently the characterization of the complete finite-size spectrum in terms of a Y-system (see the chapter \cite{chapTBA} of this review) serve that purpose.  However, a direct derivation of the Y-system from a quantum monodromy matrix is still unknown. 

A brief comment on other formulations of the superstring in $AdS_5\times S^5$ is in place. In the pure spinor string, it is possible to find a flat connection that confirms the classical integrability and an associated algebraic curve \cite{Vallilo:2003nx, Berkovits:2004jw} . It has been argued based on BRST-cohomology that the charges generated from the monodromy matrix of this flat connection exists to all orders in the $\alpha'$ expansion \cite{Berkovits:2004xu}, which has been confirmed to subleading order in \cite{Mikhailov:2007mr}. 

The outline of this part of the review is as follows: we begin in section \ref{sec:CI} by reviewing the Lax connection and monodromy matrix of the $AdS_5 \times S^5$ string. We then define the algebraic curve in terms of the quasi-momenta (which are essentially the eigenvalues of the monodromy matrix). In section \ref{sec:QuasiMom} we give a characterization of the quasi-momenta in terms of their asymptotics, poles structure etc. The example of the circular string in $S^3$ is rephrased in terms of the algebraic curve in section \ref{sec:Circular}. 
 In section \ref{sec:SYMCurve} we briefly discuss the algebraic curve of the dual $\mathcal{N}=4$ SYM theory at one-loop. In section \ref{sec:SemiClassics} the general procedure for the semi-classical quantization is presented, and a general expression for the one-loop energy shift is derived from the algebraic curve.  We furthermore show, that from this general analysis it is straightforward to compute the energy shift for the circular string of section \ref{sec:Circular}.

 Relation to other parts of the review: \\
 The relevant superstring action for the $AdS_5 \times S^5$ string was described in \cite{chapSpinning}. The Lax connection and  monodromy matrix  were already introduced in \cite{chapSigma}. Classical finite-gap solutions and their semi-classical quantization from the sigma-model point of view was discussed in \cite{chapQString}. The present part of the review gives an alternative point of view on the material in \cite{chapQString}, which manifestly relies on the integrable structure of the theory. 

%%%%%%%%%%%%%%%%%%%%%%%%%%%%%%%%%%%%%%%%%
%%%%%%%%%%%%%%%%%%%%%%%%%%%%%%%%%%%%%%%%%

\section{Classical Integrability and Spectral Curve}
\label{sec:CI}
%%%%%%%%%%%%%%%%%%%%%%%%%%%%%%%%%%%%%%%%%

\subsection{Lax connection and monodromy matrix for \texorpdfstring{$AdS_5 \times S^5$}{AdS5xS5}}

Recall that a classical sigma-model is integrable if its equation of motion can be put into zero-curvature form, with a Lax connection $L_\alpha({\bf \sigma}, {\bf \tau}, z)$ depending on the spectral parameter $z$, where $\alpha = \sigma, \tau$ denotes the world-sheet coordinates:
\begin{equation}
\partial_\alpha L_\beta - \partial_\beta L_\alpha - [ L_\alpha, L_\beta] =0 \,. 
\end{equation}
{}From the Lax connection we can form the monodromy matrix, by parallel transport along the ${\bf \sigma}$ direction of the world-sheet, along some path $\gamma$
\begin{equation}\label{MonoDef}
\Omega (z) = \mathcal{P} \exp \left(\int_0^{2\pi} L_\sigma ({\bf \sigma}, {\bf \tau}, z ) \right) \,.
\end{equation}

The classical superstring on $AdS_5 \times S^5$ is described in terms of the Green-Schwarz action by Metsaev and Tseytlin (see also \cite{chapSpinning}) as a sigma-model into the supercoset space
\begin{equation}
\frac{PSU(2,2|4)}{SO(4,1) \times SO(5)} \supset AdS_5 \times S^5  \,.
\end{equation}
A useful description of the superstring action is in terms of the supercurrents for the map from the world-sheet into the supergroup $g: \Sigma  \rightarrow PSU(2,2|4)$ which is gauged by the left-action
\begin{equation}
g \rightarrow g H\,, \qquad H \in  SO(4,1) \times SO(5) \,.
\end{equation}
Define the currents as 
\begin{equation}
J = - g^{-1} \hbox{d} g  \in \mathfrak{psu}(2,2|4) \,,
\end{equation}
which is flat $d J - J\wedge J =0$ and transforms as $J \rightarrow H^{-1} J H$\,.

The superalgebra $\mathfrak{psu} (2,2|4)$ has a $\mathbb{Z}_4$ grading 
\begin{equation}
\mathfrak{psu} (2,2|4) = \mathfrak{g}^{(0)} \oplus  \mathfrak{g}^{(1)} \oplus  \mathfrak{g}^{(2)} \oplus  \mathfrak{g}^{(3)} \,, 
\end{equation}
and we shall decompose the currents accordingly as
\begin{equation}
J = J^{(0)} + J^{(1)} + J^{(2)} + J^{(3)} \,.
\end{equation}
The action for the superstring in $AdS_5\times S^5$ then takes the form
\begin{equation}\label{AdS5S5Action}
S = {\sqrt{\lambda} \over 4 \pi} \int \hbox{STr} \left( J^{(2)} \wedge \ast J^{(2)} - J^{(1)} \wedge J^{(3)} + \Lambda \wedge J^{(2)}\right) \,, 
\end{equation}
where the Lagrange multiplier $\Lambda$ in the last term ensures the super-tracelessness of $J^{(2)}$, as is required for $PSU(2,2|4)$.

%%%%%%%%%%%%%%%%%%%%%%%%%%%%%%%%%%%%%%%%%

\subsection{Spectral Curves: Generalities}

Before discussing the curve for $AdS_5 \times S^5$ we should first elaborate on spectral curves for classical integrable systems more generally, and point out important aspects. Consider a classical integrable system, described by a Lax connection $L(x)$ and monodromy matrix $\Omega(x)$. 
The spectral curve is a complex curve defined defined by the eigenvalue equation for $\Omega(x)$
\begin{equation}\label{SpecCurve}
\hbox{SDet} \left(y \, \hbox{Id}- \Omega(x) \right) = 0 \,.
\end{equation}
It is generically not an algebraic curve and may have essential singularities and infinite genus. 
A useful subclass of configurations, the so-called ``finite gap" solutions, are such that the spectral curve is of finite genus, in this instance referred to then as ``algebraic curve".
These curves may still have singular points, which however can be desingularized by standard algebraic geometric methods, e.g.\ by small resolutions, and we shall now distinguish these two birationally equivalent curves in the following. Naturally, the curve defined by (\ref{SpecCurve}) can be written in terms of the eigenvalues $\lambda_i(x)$ of $\Omega(x)$. However, these will exhibit essential singularities in the spectral parameter, and it is more convenient to study the so-called quasi-momenta, $p_i$, where $\lambda_i(x) = e^{i p_i(x)}$. In what follows, we shall specify the curve entirely in terms of the properties of the quasi-momenta. 
   For more details on e.g.\ the maps between the various descriptions, see \cite{BabelonBernard, Vicedo:2008jk}.

%%%%%%%%%%%%%%%%%%%%%%%%%%%%%%%%%%%%%%%%%

\subsection{Algebraic Curve for $AdS_5 \times S^5$}

In \cite{Bena:2003wd} it was demonstrated that the classical equations of motion for this action are equivalent to the flatness of a one-parameter family of connections (Lax connection), thus establishing the classical integrability of the theory. The Lax connection depends on the spectral parameter, which will be denoted by $x \in \mathbb{C}$ and is given as 
\begin{equation}
L(x) = J^{(0)} + {x^2 + 1 \over x^2 -1} J^{(2)} - {2x \over x^2 -1} (\ast J^{(2)} - \Lambda)+ \sqrt{x+1 \over x-1} \, J^{(1)} + \sqrt{x-1 \over x+1} \, J^{(3)} \,.
\end{equation}
For all $x$ this is a flat connection $dL (x) - L(x)\wedge L(x) =0$.
As in (\ref{MonoDef}) we can define the corresponding monodromy matrix by parallel transport along a closed path $\gamma$, encircling the compact world-sheet direction
\begin{equation}
\Omega (x) = \mathcal{P} \exp \left( \int_\gamma L(x)\right) \,.
\end{equation}
Super-tracelessness of $L(x)$ implies unimodularity  SDet$\Omega(x) =1$. We can diagonalize $\Omega(x)$ and denote the eigenvalues by $e^{i p(x)}$, where $p(x)$ are the {\it quasi-momenta}. More specifically, we obtain
\begin{equation}  
\Omega (x) \sim \hbox{Diag} \left(e^{i \hat{p}_1 (x)}\,, \,e^{i \hat{p}_2 (x)}\,, \, e^{i \hat{p}_3 (x)}\,, \, e^{i \hat{p}_4 (x)}   | 
e^{i \tilde{p}_1 (x)}\,, \, e^{i \tilde{p}_2 (x)}\,, \, e^{i \tilde{p}_3 (x)}\,, \, e^{i \tilde{p}_4 (x)}\right)\,,
\end{equation}
where $\hat{p}$ denotes the eigenvalues corresponding to $AdS_5$ and $\tilde{p}$ to $S^5$. From unimodularity of $\Omega(x)$ it follows that\footnote{The Lagrange multiplier $\Lambda$, cf. (\ref{AdS5S5Action}), would correspond to an unphysical, overall shift, and will be ignored from now on \cite{Beisert:2005bm}.}.
\begin{equation}\label{CurveEq}
\sum_{i=1}^4 \hat{p}_i (x) - \tilde{p}_i (x)  = 2 \pi k \,,\qquad k\in \mathbb{Z} \,.
\end{equation} 
By definition, the eigenvalues $e^{i p(x)}$ are the zeroes of the characteristic polynomial of $\Omega(x)$, and as we shall define in the next section, the quasi-momenta $p(x)$ define the spectral curve. 
More precisely, the equation (\ref{CurveEq})  entails that $p$ is a multivalued function of $x$, or alternatively, it is a single-valued function of a cover of the complex $x$-plane, which defines the spectral curve. In the next section we will give a characterization of the quasi-momenta and of the resulting curve. 
The degree of the characteristic polynomial specifies the number of sheets of the cover, which in the case of the $AdS_5\times S^5$ string is eight. 

The key insight of \cite{Kazakov:2004qf} was that classical solutions can be equivalently characterized in terms of this algebraic curve, or alternatively, the quasi-momenta.

%%%%%%%%%%%%%%%%%%%%%%%%%%%%%%%%%%%%%%%%%

\subsection{Characterization of Solutions by Quasi-momenta}
\label{sec:QuasiMom}

In this section we will give a hands-on description of how classical solutions are encoded in terms of the quasi-momenta. 
This will be exemplified in the next subsection. 

Classical solutions with global conserved charges $(S_1, S_2, J_1, J_2, J_3)$ and energy $E$ will be encoded in terms of quasi-momenta.  Here $(E, S_1,S_2)$ labels weights of the $SO(4,2)$ and $(J_1, J_2, J_3)$ of the $SO(6)$ isometry groups of $AdS_5\times S^5$. 
Rather than solving an equivalent of the classical equations of motion, we lay out constraints, that will fully characterize the quasi-momenta in terms 
of asymptotics (which will be fixed by the global charges), behaviour at poles (which arise due to the pole in the Lax connection), symmetries (from the automorphism of the Lie-superalgebra $\mathfrak{psu} (2,2|4)$), and finally the so-called filling fractions. 
We will now discuss all these points in detail:
\begin{itemize}

\item The eight sheets are connected by cuts. Each of these connects two sheets, e.g. $i$ and $j$, and will be denoted by $\mathcal{C}^{ij}$. The quasi-momenta will have discontinuities along these branch-cuts 
\begin{equation}\label{CurveDef}
p_i(x+ i \epsilon) - p_j (x - i \epsilon) = 2 \pi n_{ij} \,,\qquad x \in \mathcal{C}^{ij}_n
\end{equation}
for the combination of sheets
\begin{equation}
i \in \{\tilde{1}, \tilde{2}, \hat{2}, \hat{2} \}\,, \qquad 
j \in \{\tilde{3}, \tilde{4}, \hat{3}, \hat{4} \} \,.
\end{equation}
Note that these cuts arise from the diagonalization of $\Omega$, and are thus intrinsic to the spectral data. 
The classical curve only depends on the branch-points, however, in the quantum theory, the cuts become meaningful. 
This will become clear, in the section on spin-chain spectral curves, where the cuts are shown to be condensates of Bethe roots. 

More specifically, we can associate with cuts stretching between sheets of various types a "polarization".  These correspond precisely to the sixteen physical polarization of the superstring in $AdS_5\times S^5$ and are identified in the algebraic curve in terms of cuts connecting the following pairs of sheets:
\begin{equation}\label{Pols}
\begin{aligned}
  S^5 :& \quad (\tilde{1},\tilde{3})\,,(\tilde{1},\tilde{4})\,,(\tilde{2},\tilde{3})\,,(\tilde{2},\tilde{4}) \cr
AdS_5 :& \quad (\hat{1},\hat{3})\,,(\hat{1},\hat{4})\,,(\hat{2},\hat{3})\,,(\hat{2},\hat{4}) \cr
\text{Fermions}
      :& \quad (\tilde{1},\hat{3})\,,(\tilde{1},\hat{4})\,,(\tilde{2},\hat{3})\,,(\tilde{2},\hat{4})\cr
       & \quad (\hat{1},\tilde{3})\,,(\hat{1},\tilde{4})\,,(\hat{2},\tilde{3})\,,(\hat{2},\tilde{4}) \,.
\end{aligned}
\end{equation}
The situation is depicted in figure 1, where both macroscopic cuts, that correspond to a classical solution are shown, as well as all the physical excitations from (\ref{Pols}).

%%%%%%%%%%%%%%%%%%%%%%%%%%%%%%%
\begin{figure}
\begin{center} \label{fig:CurveSheets}
  \includegraphics*[width =12cm]{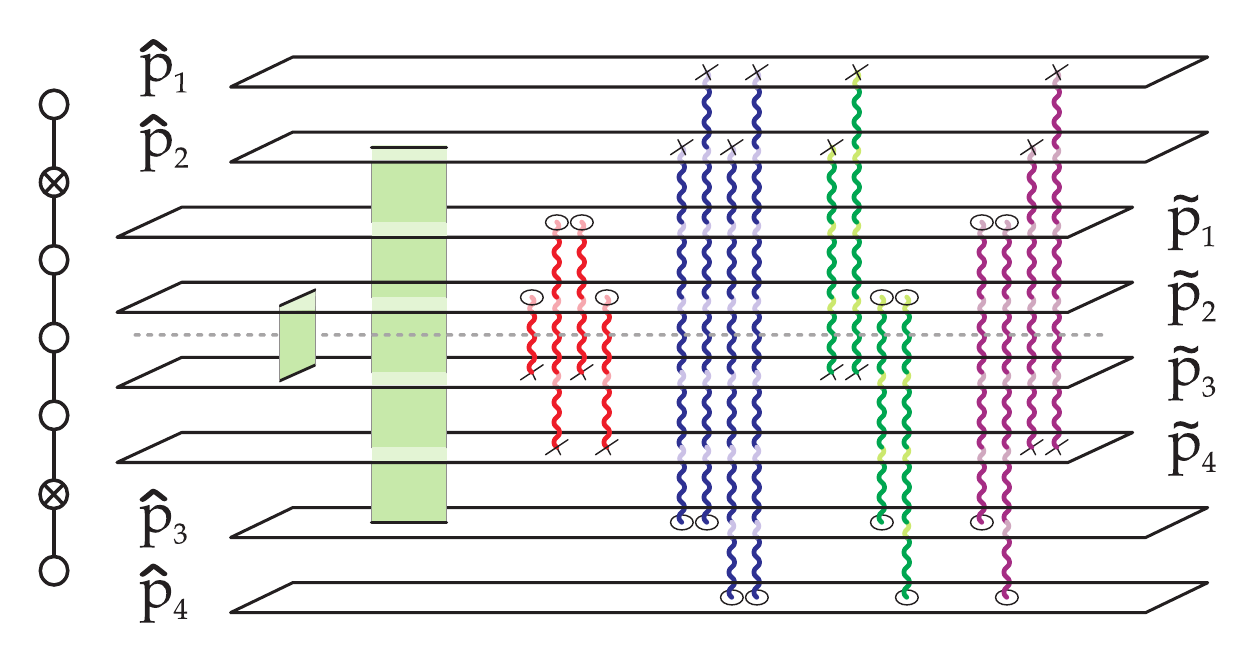}  
  \caption{The spectral curve for classical superstrings on $AdS_5 \times S^5$. The sheets are connected by cuts (green), which characterize classical solutions. The left most cut alone, e.g. corresponds to a one-cut solution in the $S^3 \times \mathbb{R}$ subspace, whereas the second cut is supported in $AdS_3 \times S^1$. 
  The remaining part of the graph depicts all polarization of physical fluctuations.  Red:  bosonic fluctuations in the $S^5$ direction. 
  Blue: bosonic fluctuations in the $AdS_5$ direction. Green and purple: fermionic fluctuations. }
  \end{center}
\end{figure}
%%%%%%%%%%%%%%%%%%%%%%%%%%%%%%%%%%%

\item The quasi-momenta  have  poles in the $x$-plane at $x =\pm 1$ -- which can be readily seen from the Lax connection, which has poles at $x= \pm 1$ --  with residues that are correlated due to the Virasoro constraint
\begin{equation}\label{VirConst}
\{\hat{p}_1,\hat{p}_2,\hat{p}_3,\hat{p}_4|\tilde{p}_1,\tilde{p}_2, \tilde{p}_3, \tilde{p}_4\}
=
\frac{\{  \alpha_\pm,  \alpha_\pm,  \beta_\pm,  \beta_\pm|  \alpha_\pm,  \alpha_\pm,  \beta_\pm,  \beta_\pm\}}{x\pm 1}  + O(1)\,.
\end{equation}

\item Global charges of the classical solution determine the asymptotics of the quasi-momenta for $x\rightarrow \infty$. This follows simply from the fact that in this limit, the Lax connection $L(x)$ reduces to the Noether current. 
It is useful to rescale the global $\mathfrak{psu} (2,2|4)$ charges by $1/\sqrt{\lambda}$ and define $\mathcal{Q} = Q/\sqrt{\lambda}$. Then the asymptotics of the quasi-momenta are
\begin{equation} \label{AsympInfty}
\left( 
\begin{array}{c}
\hat{p}_1\\
\hat{p}_2\\
\hat{p}_3\\
\hat{p}_4\\   \hline
\tilde{p}_1\\
\tilde{p}_2\\
\tilde{p}_3\\
\tilde{p}_4\\
\end{array}\right) = \frac{2\pi }{x}
\left(\begin{array}{l}
+\mathcal{E}-\mathcal{S}_1+\mathcal{S}_2 \\
+\mathcal{E}+\mathcal{S}_1 -\mathcal{S}_2 \\
-\mathcal{E}-\mathcal{S}_1  -\mathcal{S}_2 \\
-\mathcal{E}+\mathcal{S}_1 +\mathcal{S}_2 \\ \hline
+\mathcal{J}_1+\mathcal{J}_2-\mathcal{J}_3  \\
+\mathcal{J}_1-\mathcal{J}_2+\mathcal{J}_3 \\
-\mathcal{J}_1+\mathcal{J}_2 +\mathcal{J}_3 \\
-\mathcal{J}_1-\mathcal{J}_2-\mathcal{J}_3
\end{array}\right)   + O\left({1 \over x^2}\right)\,.
\end{equation}
For the spectral problem it is in particular of interest to note that the energy $\mathcal{E} = E/\sqrt{\lambda}$ can be extracted from these asymptotics
\begin{equation}
E = {\sqrt{\lambda}\over 4 \pi} \lim_{x\rightarrow \infty} x (\hat{p}_1(x) + \hat{p}_2 (x)) \,.
\end{equation}
We will see later, how this is done in practice. 

\item The quasi-momenta are furthermore restricted by an automorphism of the algebra $\mathfrak{psu}(2,2|4)$, which  imposes the following relations for the quasi-momenta
\begin{equation}\label{Auto}
\begin{aligned}
\tilde{p}_{1,2}(x)&=-\tilde{p}_{2,1}(1/x)-2\pi m\cr
\tilde{p}_{3,4}(x)&=-\tilde{p}_{4,3}(1/x)+2\pi m\cr
\hat{p}_{1,2,3,4}(x)&=-\hat{p}_{2,1,4,3}(1/x)\,.
\end{aligned}
\end{equation}
This inversion symmetry allows us to determine the quasi-momenta inside the region $|x|< 1$. 

\item Finally, for each cut, we define the filling fraction 
\begin{equation}\label{FillFrac}
S_{ij}=\pm\, {\sqrt{\lambda}\over 8\pi^2i} \oint_{\mathcal{C}_{ij}} \left(1-{1 \over x^2}\right) p_i(x) dx\,.
\end{equation}
These are the action angle variables for the theory \cite{Dorey:2006mx}. These curve data specify precisely a macroscopic excitations of the string with $S_{ij}$ quanta of mode number $n$. 

\end{itemize}

%%%%%%%%%%%%%%%%%%%%%%%%%%%%%%%%%%%%%%%%%

\subsection{Example: Circular String}
\label{sec:Circular}

To illustrate the spectral curve method, we now describe the circular string solution with support in $S^3 \times \mathbb{R}$ \cite{Frolov:2003qc, Gromov:2007aq}. We restrict to the case, when all $\mathfrak{su}(4)$  spins $J_i$ are equal, and parametrize the solution by one spin $J = \sqrt{\lambda} \mathcal{J}$. We furthermore restrict to the case of a single cut. 
Since this solution has trivial support in the $AdS_5$ direction, the corresponding quasi-momenta are determined simply in terms of trivial asymptotics at infinity and the correct pole structure at $\pm 1$. The poles are correlated as required by (\ref{VirConst}) and determine the quasi-momenta as
\begin{equation}
\hat{p}_1 = \hat{p}_2 = - \hat{p}_3  = - \hat{p}_4 = {2 \pi \kappa x \over x^2-1} \,.
\end{equation}
The quasi-momenta associated to the $S^5$ directions will have cuts, and have to be consistent with the inversion symmetry. In \cite{Gromov:2007aq} these were determined as 
\begin{equation}\label{CircularP}
\left( 
\begin{array}{c}
\tilde{p}_1 \cr
\tilde{p}_2 \cr
\tilde{p}_3 \cr
\tilde{p}_4 
\end{array}
\right)
= \left( 
\begin{array}{l}
 {x\over x^2 -1} K(1/x) \cr
 {x\over x^2- 1} K(x) - m \cr
  {x\over 1-x^2 } K(x) + m \cr
 {x\over 1-x^2} K(1/x) 
\end{array}
\right)
\end{equation}
where $K(x) = \sqrt{m^2 x^2 + \mathcal{J}}$. The cut extends along the imaginary axis and from the various constraints
\begin{equation}
\mathcal{E} = \kappa = \sqrt{\mathcal{J}^2 + m^2} \,.
\end{equation}
It is in general not so easy to reverse-engineer the solution from the quasi-momenta. However, for many aspects, in particular computing the spectrum, it is a particularly powerful way to describe solutions.

%%%%%%%%%%%%%%%%%%%%%%%%%%%%%%%%%%%%%%%%%

\section{The Algebraic Curve of \texorpdfstring{$\mathcal{N}=4$}{N=4} SYM}
\label{sec:SYMCurve}

So far our discussion of the spectral curve was focused on the classical $AdS_5 \times S^5$ string. However, there is a spectral curve also for the dual $\mathcal{N}=4$ SYM theory. At one-loop it was shown that the eigenvalues of the dilatation operator can be equivalently computed from a ferro-magnetic Heisenberg spin chain with $\mathfrak{psu} (2,2|4)$ symmetry, which can be diagonalized using a Bethe ansatz \cite{Minahan:2002ve, Beisert:2004ry, Beisert:2003jj, Beisert:2003yb, Beisert:2005fw}. In the thermodynamic limit Bethe roots condense and form cuts. The resulting structure is precisely an algebraic curve, which intriguingly resembles the curve for the superstring  \cite{Kazakov:2004nh, Beisert:2004ag, SchaferNameki:2004ik, Beisert:2005di, Beisert:2005bm}. We will now briefly summarize the construction of the SYM curve. 
For details of the Bethe ansatz see the other contributions \cite{chapChain, chapABA}.

%%%%%%%%%%%%%%%%%%%%%%%%%%%%%%%%%%%%%%%%%

\subsection{Bethe Ansatz Equations}

The one-loop dilatation operator can be diagonalized by a Bethe ansatz for a super-spin chain with symmetry $\mathfrak{psu}(2,2|4)$ and ${\bf 4} | {\bf 4}$ representation at each spin-chain site \cite{Beisert:2003yb}. 
The Bethe roots are $u_i^{(k)}$, $k=1,\cdots, r=7$ and
$i=1,\cdots, J_k$, where $J_k$ denotes the excitation
number for the $k$th root.  
Further, define $J=\sum J_k$ as the total excitation number,
$L$ be the length of the
spin chain, and denote the Cartan matrix\footnote{This is modulo signs that are discussed in \cite{chapABA}.} of $\mathfrak{psu}(2,2|4)$ by $M$ and the weight of the representation by $V$. 
Then the Bethe Ansatz equations for the nearest neighbour spin-chain are
\begin{equation}\label{BAE}
\left({u_i^{(k)} - {i\over 2} V_{k}\over u_i^{(k)} + {i\over 2} V_{k}} \right)^L
= \prod_{l=1}^{r}\prod_{j=1}^{J_l}  {u_i^{(k)} - u_j^{(l)}
-{i\over 2}M_{k l}\over  u_i^{(k)} - u_j^{(l)} +{i\over 2}M_{k l}}\,.
\end{equation}
Translational invariance along the spin chain implies further that
\begin{equation}
1 = \prod_{k=1}^r 
\prod_{i=1}^{J_k} {u_i^{(k)} + {i\over 2} V_{k} \over u_i^{(k)} -
{i\over 2} V_{k}} = e^{ i P} \,,
\end{equation}
where $P $ is the total momentum. Solving these algebraic equations for the Bethe roots determines the values of the conserved charges, in particular the energy of the spin-chain Hamiltonian, and thus the Dilatation operator at one-loop
\begin{equation}
Q_n= {i\over n-1} \sum_{l=1}^n \sum_{j=1}^{J_n} 
\left(
{1\over (u_j^{(l)} + {i\over 2} V_l)^{n-1}} - 
{1\over (u_j^{(l)} - {i\over 2} V_l)^{n-1}}  
\right)\,.
\end{equation}
In particular, the energy $E$ of the state is read off from $Q_2$ as 
\begin{equation}
 E=c g^2 Q_2 \,,
\end{equation}
for some constant $c$. 

%%%%%%%%%%%%%%%%%%%%%%%%%%%%%%%%%%%%%%%%%

\subsection{Thermodynamic Limit and Algebraic Curve}

As in the case of the superstring, the main interest is in determining the values of $Q_r$, and not in solving an auxiliary set of equations -- the classical equations of motion in the case of the superstring, or the Bethe ansatz equations in the SYM theory. 
There is an analog of the spectral curve in the SYM that arises in the limit of large number of Bethe roots. 
More precisely, the  algebraic curve of the above system arises in the thermodynamic limit $L\rightarrow \infty$.
Taking the logarithm of (\ref{BAE}) yields
\begin{equation}
L \log \left({u_i^{(k)} - {i\over 2} V_{k}\over u_i^{(k)} + {i\over 2} V_{k}} \right)
= \sum_{l=1}^r \sum_{j=1,\, j\not= i}^{J_l} \log \left(
{u_i^{(k)} - u_j^{(l)} -{i\over 2}M_{k l}
\over  u_i^{(k)} - u_j^{(l)} +{i\over 2}M_{k l}}
\right) - 2\pi i n_i^{(k)}\,,
\end{equation}
where $n_i^{(k)}\in \mathbb{Z}$ are the mode numbers, arising due to taking the logarithm. 
We now rescale the Bethe roots by $1/L$ to $x_i^{(k)} = u_i^{(k)}/L$ and take $L, J \rightarrow \infty$, while keeping $n_i^{(k)}$ fixed
\begin{equation}
- {V_{k}\over x_i^{(k)}} = \sum_{l=1}^r {1\over J_l}\sum_{j=1,\,  j\not=
i}^{J_l} {M_{k l}\over
x_i^{(k)}-x_j^{(l)}} - 2\pi n_i^{(k)} \,.
\end{equation}
It is useful to introduce a density of Bethe roots and a resolvent for their distribution
\begin{equation}
\begin{aligned}
\rho_k(x) &= 
%%{1\over J_k} 
\sum_{j=1}^{J_k} \delta \left(x-x_j^{(k)}\right)\cr
G_k(x) & = {1\over J_k} \sum_{j=1}^{J_k} {1 \over x-x_j^{(k)}} \,.
\end{aligned}
\end{equation}
In the limit, the Bethe roots condense into cuts $\mathcal{C}_k$, and the Bethe equations take the continuum form
\begin{equation}
\slashint_{\mathcal{C}} dv \, {\rho_k(v) M_{k f(v)}\over v-u} = -{V_{k}\over u} + 2\pi n_i^{(k)} \,, \qquad u\in \mathcal{C}_i^{(k)}\,,
\end{equation}
where $\mathcal{C}=\cup_k \mathcal{C}_k$ and each of the curves $\mathcal{C}_k$ associated to
simple roots is on the other hand $\mathcal{C}_k=\cup_j \mathcal{C}_j^{(k)}$.
% Further, $f(u)= l$ for $u\in\mathcal{C}_l\subset\mathcal{C}$.
This can equivalently be written in terms of the resolvent $G_k(u) $ in the continuum limit
\begin{equation}
M_{k k} \Gslash_k(u)+  \sum_{l\not= k}  M_{k l} G_l(u) =
-{V_{k} \over u} + 2\pi n_i^{(k)}  \,,\qquad u \in \mathcal{C}_i^{(k)} \,. 
\end{equation}
Slashes denote principal values. This equation can be put into a more familiar form by writing them in terms of the singular resolvents $\tilde{G}$, where the  poles in $1/u$ have been absorbed into the definition of the resolvent, and furthermore taking linear combinations (the quasi-momenta) $p_i \sim \pm (\tilde{G}_{i-1} - \tilde{G}_i) $ (for details see \cite{SchaferNameki:2004ik}) so that we arrive at
\begin{equation}
M_{kk}\tilde{\Gslash}_{k} +\sum_{j\not=k} M_{kj} \tilde{G}_j(u) = \pslash_k(u) -\pslash_{k+1}(u) = 2 \pi n_j^{(k)}\,,
 \qquad u\in \mathcal{C}_j^{(k)} \,.
\end{equation}
This is precisely the type of equation that characterizes the spectral curve in the superstring case. 
Again, the asymptotics of the resolvent/quasi-momenta encode the relation to the global charges
\begin{equation}
G_k(u) = -{1\over u} \int_{\mathcal{C}_k} dv \rho_k(v) + O\left({1\over u^2}\right) 
= -{J_k \over u} + O\left({1\over u^2}\right) \,.
\end{equation}
A precise comparison of the SYM curve \cite{SchaferNameki:2004ik, Beisert:2005di} and string curve \cite{Beisert:2005bm} can be found in \cite{Beisert:2005di}. The main features are, that the asymptotics and constraints on the quasi-momenta agree up to a redefinition of the spectral parameter and modulo pole structure, and thus, also the algebraic curves are in agreement.

%%%%%%%%%%%%%%%%%%%%%%%%%%%%%%%%%%%%%%%%%
%%%%%%%%%%%%%%%%%%%%%%%%%%%%%%%%%%%%%%%%%

\section{Semi-classical Quantization of the Spectral Curve}
\label{sec:SemiClassics}

Apart from giving a general, concise description of classical solutions, the spectral curve is a powerful means to compute quantum fluctuations. In part \cite{chapQString} of the review, the quantization around classical solutions with large spins, was already described from the point of view of the sigma-model:  a classical field configuration is perturbed and the fluctuations quadratically quantized. The sum of the fluctuation frequencies make up the energy shift at one-loop (in $\alpha'$, or equivalently $1/\sqrt{\lambda}$). We will not give an alternative approach, based on the algebraic curve, and present a general expression for the one-loop shift for general solutions.

%%%%%%%%%%%%%%%%%%%%%%%%%%%%%%%%%%%%%%%%%
\subsection{Perturbing the Spectral Curve}

A classical configuration can be viewed as a continuous collection of poles which have condensed into the cuts $\mathcal{C}^{ij}$. This intuition is particularly transparent in the comparison with the algebraic curve of the Yang-Mill theory, as discussed in section \ref{sec:SYMCurve}, where indeed, the cuts arose from condensation of Bethe roots. 
{}From this point of view, semi-classical quantization naturally corresponds to adding small fluctuations, or poles, to the classical configuration. 
Naturally, these fluctuations will have polarizations, labeled by $(ij)$, and amount to shifting the 
quasi-momenta
\begin{equation}
p_i(x) \rightarrow p_i(x) + \delta^{ij} p_i(x) \,.
\end{equation}
The energy shift is then obtained as the sum over all fluctuation frequencies. 
The shifts in the quasi-momenta $\delta^{ij} p_i(x)$ are constrained by the asymptotics etc of the quasi-momenta, outlined in section \ref{sec:QuasiMom}:

\begin{itemize}
\item The perturbed quasi-momenta will have to continue to satisfy the relation \\
(\ref{CurveDef}). First we need to determine the position of the new pole $x_n^{ij}$
\begin{equation}
p_i (x_n^{ij}) - p_j (x_n^{ij}) = 2 \pi n_{ij} \,.
\end{equation}
The physical poles correspond to solutions of this equation with $|x_n^{ij}|>1$.\footnote{
The inversion symmetry maps the region $|x|>1$ maps to $|x|<1$, so that considering one of these regions (the physical region) is sufficient to describe the curve. Without loss of generality the region $|x|>1$ is chosen to be the physical region.}
The fluctuation $\delta_n^{ij} p_{i}$ will have to add a pole at $x_n^{ij}$ with residue, $\alpha(x_n^{ij})$, such that it changes the filling fraction $S_{ij}$ (\ref{FillFrac}) by one, i.e.
\begin{equation}
\delta_n^{ij} p_{i} = \pm {\alpha(x_n^{ij})\over x - x_n^{ij}} \,,
\end{equation}
with 
\begin{equation}
\alpha (x) = {4\pi \over \sqrt{\lambda}} {x^2 \over x^2-1} \,.
\end{equation}
The total shifted quasi-momentum is obtained by summing over all fluctuations with all relevant polarizations in (\ref{Pols})
\begin{equation}
\delta p_i \sim \sum_{(ij)} \delta^{(ij)} p_{i} (x) 
            = \sum_{(ij)}\epsilon_i N_n^{ij} {\alpha (x_n^{ij}) \over x- x_n^{ij}} \,,
\end{equation}
where $N_{n}^{ij}$  label the excitations with mode number $n$ and polarization $(ij)$, and the signs are 
\begin{equation}\label{FluctuSigns}
1= \epsilon_{\hat 1} = \epsilon_{\hat 2} = - \epsilon_{\hat 3} = -\epsilon_{\hat 4}
 = -\epsilon_{\tilde 1} = -\epsilon_{\tilde 2} = \epsilon_{\tilde 3} = \epsilon_{\tilde 4} \,.
\end{equation}
{}From (\ref{CurveDef}) it furthermore follows that
\begin{equation}
\delta p_i(x+ i \epsilon) - \delta p_j (x- i \epsilon) = 0 \,,\qquad x \in \mathcal{C}^{ij}_n \,.
\end{equation}

\item As in the classical case, the poles at $x= \pm 1$ have to be correlated due to the Virasoro constraint
\begin{equation}
\begin{aligned}
&\{\delta \hat{p}_1,\delta\hat{p}_2,\delta\hat{p}_3,\delta\hat{p}_4|\delta\tilde{p}_1,\delta\tilde{p}_2,\delta\tilde{p}_3,\delta\tilde{p}_4\}\cr
& \qquad \qquad =
\frac{\{  \delta\alpha_\pm,\delta  \alpha_\pm, \delta \beta_\pm, \delta \beta_\pm|
\delta\alpha_\pm, \delta \alpha_\pm, \delta \beta_\pm,  \delta\beta_\pm\}}{x\pm 1}   + O(1)\,.
\end{aligned}
\end{equation}

\item The asymptotics at infinity (\ref{AsympInfty}) of the $\delta^{ij} p_i$ can be easily read off

\begin{equation}
   \left(\begin{array}{c}
\delta \hat{p}_1\\
\delta\hat{p}_2\\
\delta\hat{p}_3\\
\delta\hat{p}_4\\  \hline
\delta\tilde{p}_1\\
\delta\tilde{p}_2\\
\delta\tilde{p}_3\\
\delta\tilde{p}_4\\
\end{array}\right)
=   {4 \pi \over x \sqrt{\lambda}  } 
\left(\begin{array}{rrl}
+\delta \Delta/2&+N_{\h 1\h 4}+N_{\h 1 \h 3}&+N_{\h 1\t 3}+N_{\h1\t 4}\\
+\delta \Delta/2&+N_{\h 2\h 3}+N_{\h 2\h 4}&+N_{\h2\t 4}+N_{\h2\t 3} \\
-\delta \Delta/2&-N_{\h 2\h 3}-N_{\h 1 \h 3}&-N_{\t 1\h3}-N_{\t 2\h3} \\
-\delta \Delta/2&-N_{\h 1\h 4}-N_{\h 2\h 4}&-N_{\t 2\h4}-N_{\t 1\h4} \\ \hline
&- N_{\t1 \t4}- N_{\t 1\t 3} &-N_{\t 1\h3}-N_{\t 1\h4} \\
&- N_{\t 2 \t3}- N_{ \t 2 \t4}                         &-N_{\t 2\h4}-N_{\t 2\h3}\\
&+ N_{ \t2 \t3}+ N_{ \t1 \t3}&+N_{\h1\t 3}+N_{\h2\t 3}\\
&+ N_{ \t1 \t4}+ N_{ \t 2 \t4}             &+N_{\h2\t 4}+N_{\h1\t 4}
\end{array}\right) + O\left({1\over x^2} \right) \,,
\end{equation}
where $\delta\Delta$ parametrizes the shift in the energy $\mathcal{E}$. 
{}From these asymptotics we can also determine the fluctuation frequencies $\Omega_{n}^{ij}$ that are familiar from the direct semi-classical quantization by
\begin{equation}\label{OmegaDef}
\Omega_n^{ij} = - 2 \delta_{i, \hat{1}} + {\sqrt{\lambda}\over 2 \pi} \lim_{x\rightarrow \infty} x \delta_n^{ij} p_{\hat{1}}(x) \,.
\end{equation}
The energy shift then takes the usual form, as sum over fluctuation frequencies
\begin{equation}
\delta \Delta = \sum_{ij,n} N_{ij}^n \Omega_n^{ij} \,.
\end{equation}

\item Finally, the inversion symmetries extend trivially to the shifted quasi-momenta. These rather inconspicuous transformations, however, turn out to be rather powerful in determining the energy shifts. We shall see in section  \ref{sec:CompleteShifts} how one can derive a closed expression for the one-loop energy shift, by invoking the asymptotics, pole structure, and inversion symmetry. 

\end{itemize}

So far we covered all the constraints that follow from the asymptotics of the classical
quasi-momenta. In addition, the fluctuations will backreact upon the classical cuts and close to the branch-points (or cut-endpoints) we impose for $p_i \sim \sqrt{(x-a)}$ close to the branch-point $x=a$
\begin{equation}
\delta p_i \sim {d \over d x} p_i \,.
\end{equation}
Solving these constraints in particular fixes $\delta E$, which is the desired one-loop energy shift.

%%%%%%%%%%%%%%%%%%%%%%%%%%%%%%%%%%%%%%%%%

\subsection{General expression of one-loop energy shift}
\label{sec:CompleteShifts}

Rather than presenting examples of computations of energy shifts using the algebraic curve, which can e.g. be found for a plentitude of solutions (BMN, spinning string solutions, giant magnon) in the literature listed in the introduction, it is perhaps more interesting to point out that using general properties of the quasi-momenta constrain the energy shift such that closed expressions can be obtained for fairly general solutions (for any number of cuts) \cite{Gromov:2008ec}. 
We then apply it to the circular string solution of section \ref{sec:Circular}. This will be essentially a trivial step, once the general energy shift has been derived, and hopefully exemplifies that the algebraic curve approach is indeed very powerful for computing these effects. 

%%%%%%%%%%%%%%%%%%%%%%%%%%%%%%%%%%%%%%%%%

\subsubsection{Off-shell Fluctuation Frequencies}

The key idea is to introduce the concept of an {\it off-shell fluctuation} (also sometimes refered to as quasi-energies), which means, defining the fluctuation as  a function of the spectral parameter $x$ and a variable $y$, such that the following holds
\begin{equation}
\delta_n^{ij} p_k(x)=\left. \delta^{ij} p_k(x;y) \right|_{y=x_n^{ij}} \,.
\end{equation}
This off-shell flucutation $ \delta^{ij} p_k(x;y)$ is fixed by the same asymptotics as the \textit{on-shell} shift of quasimomenta $\delta_n^{ij} p_k(x)$ except that the position of the pole is left unfixed. In the same way, we can then define off-shell fluctuation energies, by applying the same reasoning to (\ref{OmegaDef})
\begin{equation}
\Omega_n^{ij}=\left.\Omega^{ij}(y)\right|_{y=x_n^{ij}} \,.
\end{equation} 
The off-shell frequency is related for the particular case of the $SU(2)$ principal chiral model to the quasi-energy introduced in  \cite{Vicedo:2008jy}.

It is simple to reconstruct the off-shell frequency from a given on-shell one $\Omega_n^{ij}$. We know that the mode number $n$ is determined precisely by the requirement $p_i(x_n^{ij}) - p_j (x_n^{ij}) = 2 \pi n$, so that reverting this relation, treating $n$ now as a function of $p_l(y)$ we obtain
\begin{equation}
\Omega^{ij}(y) =\left.  \Omega_n^{ij} \right|_{n\to \frac{p_i(y)-p_j(y)}{2\pi}} \,.
\end{equation}

We will now explain how, using the inversion symmetry (\ref{Auto}), we can relate many  off-shell fluctuation energies. In this way we will find a powerful reduction algorithm for the computation of the fluctuation energies and thus the one loop energy shift
\begin{equation}\label{OneLoopE}
\delta \Delta^{1-loop}=\frac{1}{2} \sum_{ij,n}(-1)^{F_{ij}} \Omega_n^{ij} \,,
\end{equation}
around a generic classical solution.

%%%%%%%%%%%%%%%%%%%%%%%%%%%%%%%

\subsubsection{Frequencies from inversion symmetry}

An important property of the quasi-momenta, which follows from the $\mathbb{Z}_4$-grading of the $\mathfrak{psu}(2,2|4)$ superalgebra, is the inversion symmetry (\ref{Auto}) under $x\rightarrow 1/x$, which exchanges the quasi-momenta $p_{\t1,\t4} \leftrightarrow p_{\t2,\t 3}$ and likewise for the $AdS$ hatted quasi-momenta.  Thereby, a  pole connecting the sheets $(\t2, \t3)$ at position $y$, always comes with an image pole at position $1/y$ connecting the sheets $(\t1,\t4)$.
We can obtain a physical frequency  $\Omega^{\t 1\t 4}(y)$, by analytically continuing the off-shell frequency $\Omega^{\t 2 \t 3}(y)$, inside the unit circle. This is because when we cross the unit-circle, the physical pole for $(\t2\t3)$ becomes unphysical, thereby rendering its image, which lies now outside the unit-circle, a physical pole for $(\t1\t4)$.
More precisely, it was shown in \cite{Gromov:2008ec}, that with this kind of reasoning we can compute the $(\h1 \h4)$ fluctuation in terms of the $(\h2\h3)$ one. For the $AdS$ fluctuations, indeed, the relation is
\begin{equation}\label{OmegaMapAdS}
\Omega^{\h1\h4}(y)=-\Omega^{\h2 \h3}(1/y)-2 \,.
\end{equation}
This follows by invoking the general pole/asymptotics of the quasi-momenta and in the inversion symmetry. 

Similarly we can proceed for the $S^5$ frequencies and relate $\Omega^{\t2\t3}(y)$ with $\Omega^{\t1 \t4}(y)$.
It is clear that $\Omega^{\t1\t4}(y) = -  \Omega^{\t2 \t3}(1/y)$ +constant, which can be fixed from $\Omega^{\t1\t4}(\infty)=0$. Thus,  the relation is similar to (\ref{OmegaMapAdS}), except that the constant term differs:
\begin{equation}\label{InversionOmega}
\Omega^{\t1\t4}(y) = -  \Omega^{\t2 \t3}(1/y) +\Omega^{\t2 \t3}(0)\,.
\end{equation}
For the purpose of computing the one-loop shift these constants are irrelevant and can be shown to cancel in the sum over frequencies\footnote{Note, that in the case of $AdS_4 \times \mathbb{CP}^3$ these constants play an important role.}. 

So far we have obtained the frequencies $(14)$ from $(23)$. In the next subsection we will show how to derive all remaining frequencies. For a very large class of classical solutions we will be able to extract all fluctuation energies, including the fermionic ones, from the knowledge of a single $S^3$ and a single $AdS_3$ fluctuation energy.

%%%%%%%%%%%%

\subsubsection{Basis of fluctuation energies}

For simplicity we consier only symmetric classical configurations that have pairwise symmetric quasi-momenta
\begin{equation}\label{SymSol}
p_{\h1, \h 2 ,\t 1, \t2 } = - p_{\h4, \h3, \t 4 , \t3} \,,
\end{equation}
This is in particular the case for all rank one solutions, i.e. $\mathfrak{su}(2)$ and $\mathfrak{sl}(2)$, however, a generalization to other cases should not be difficult. 

Consider e.g. the fermionic frequency $\Omega^{\h2 \t3}(y)$. This energy can be thought of as a linear combination of the physical fluctuation $\Omega^{\t 2 \t 3}(y)$ and an unphysical fluctuation $\Omega^{\h2 \t2}(y)$ (it is unphysical, as it is not one of the fluctuations in (\ref{Pols}))  momentum-carrying polarisations
\begin{equation}\label{FermLinearCombi}
\Omega^{\h2 \t3}(y) =  \Omega^{\t 2 \t 3}(y)+ \Omega^{\h2 \t2}(y) \,.
\end{equation}
Since we are considering symmetric configurations, this unphysical fluctuation energy is identical to $\Omega^{\t3\h3}(y)$, i.e.
\begin{equation}
 \Omega^{\h2 \t2}(y) = \Omega^{\t3\h3}(y) \,.
\end{equation}
As in (\ref{FermLinearCombi}), these unphysical fluctuations can be linearly combined in terms of physical fluctuations
\begin{equation}
\Omega^{\h2 \h3}(y)  =\Omega^{\h2 \t2}(y) + \Omega^{\t2 \t3}(y)+ \Omega^{\t3 \h3}(y) \,.
\end{equation}
Combining all these relations we obtain
\begin{equation}\label{LinearCombiExample}
\Omega^{\h2 \t3}(y) = {1\over 2} \left(  \Omega^{\t 2 \t 3}(y) + \Omega^{\h2 \h3}(y)   \right) \,.
\end{equation}
Proceeding in a similar fashion all frequencies can be obtained as linear combinations of $\Omega^{\t 2 \t 3}(y)$ and
$\Omega^{\h2 \h3}(y)$. 

%%%%%%%%%%%%%%%%%%%%%%%%%%%%%%%%%%%%%%%%%
%%%%%%%%%%%%%%%%%%%%%%%%%%%%%%%%%%%%%%%%%

\subsubsection{Final result}

The physical frequencies are labeled by the eight bosonic and eight fermionic polarizations (\ref{Pols}), so we can label them by
\begin{equation}
\Omega^{ij} \,,\qquad \hbox{where}\quad   i= (\h1, \h2, \t1, \t 2) \qquad   j = (\h3 ,\h4,  \t3 ,\t4) \,.
\end{equation}
To construct the complete set of off-shell frequencies for a symmetric solution (\ref{SymSol})
in terms of the two fundamental $S^3$ and $AdS_3$ ones $\Omega^{\t2 \t3}(y)$ and $\Omega^{\h2 \h3}(y)$ and their images under $y\rightarrow 1/y$, we first construct by inversion
\begin{equation}
\begin{aligned}\label{FreqTable1}
\Omega^{\t1\t4} (y) &= -  \Omega^{\t2 \t3}(1/y) + \Omega^{\t2 \t3}(0) \cr
\Omega^{\h1\h4} (y) & = - \Omega^{\h2 \h3}(1/y) -2  \,.
\end{aligned}
\end{equation}
The remaining frequencies are then obtained by linear combination of these four fluctuation frequencies. In this way we obtain the following concise form for all off-shell frequencies
\begin{equation}\label{OmegaFinal}
\Omega^{ij} (y) = {1\over 2} \left(\Omega^{ii'} (y) + \Omega^{j'j} (y) \right) \,,
\end{equation}
where
\begin{equation}
 (\h1, \h2, \t1, \t 2,\h3 ,\h4,  \t3 ,\t4)' = (\h4, \h3, \t 4 , \t 3 ,\h 2, \h 1 , \t 2, \t 1) \,.
\end{equation}
To finally, make the point, that these are indeed written in terms of the basis  frequencies $\Omega^{\t2 \t3} (y)$ and $\Omega^{\h2 \h3}(y)$, we present the complete set of frequencies
\begin{equation}
\begin{aligned}\label{FreqTable}
\Omega^{\t1\t4} (y) &= -  \Omega^{\t2 \t3}(1/y) +\Omega^{\t2\t3}(0)\cr 
\Omega^{\t2 \t4} (y) =
\Omega^{\t1\t3}(y)  &= {1\over 2} \left( \Omega^{\t2 \t3}(y) + \Omega^{\t1\t4} (y) \right)
                               = {1\over 2} \left( \Omega^{\t2 \t3}(y) - \Omega^{\t2\t3} (1/y)+\Omega^{\t2\t3}(0) \right)
                                 \cr
 \Omega^{\h1\h4} (y) & = - \Omega^{\h2 \h3}(1/y) -2 \cr
\Omega^{\h2 \h4} (y) =
\Omega^{\h1 \h3} (y) &= {1\over 2}  \left( \Omega^{\h2 \h3}(y) + \Omega^{\h1\h4} (y) \right)
				    = {1\over 2}  \left( \Omega^{\h2 \h3}(y) - \Omega^{\h2\h3} (1/y) \right) -1 \cr
\Omega^{\h2\t4}(y) =
\Omega^{\t1 \h3} (y) &= {1\over 2} \left(\Omega^{\h2 \h3}(y)+ \Omega^{\t1\t4} (y)  \right)
				   = {1\over 2} \left( \Omega^{\h2 \h3}(y)- \Omega^{\t2 \t3}(1/y)+\Omega^{\t2\t3}(0) \right)  \cr				
\Omega^{\t2\h4}(y)=
\Omega^{\h1 \t3} (y) &= {1\over 2} \left( \Omega^{\t2\t3} (y) + \Omega^{\h1 \h4}(y)\right)
				 = {1\over 2}  \left( \Omega^{\t2\t3} (y) - \Omega^{\h2 \h3}(1/y)\right)   -1 \cr
\Omega^{\t1 \h4} (y) =
\Omega^{\h1 \t4} (y) &= {1\over 2} \left( \Omega^{\t1 \t4}(y) + \Omega^{\h1\h4} (y) \right)
				   = {1\over 2} \left(- \Omega^{\t2\t3}(1/y) - \Omega^{\h2\h3}(1/y)+\Omega^{\t2\t3}(0) \right)
				   -1  \cr	
\Omega^{\h2\h3}(y)=
\Omega^{\t2 \h3} (y) &= {1\over 2} \left( \Omega^{\t2 \t3}(y) + \Omega^{\h2\h3} (y) \right)	 \,.
\end{aligned}
\end{equation}

%\begin{figure}
%\begin{center}
%  \includegraphics*[width =14cm]{fermion.eps}  \caption{\small
%  Depiction of equation (\ref{LinearCombiExample}). Physical excitations are coloured in red/blue/green. The unphysical intermediate frequency is depicted in grey. \label{fermion}
% }
%  \end{center}
%\end{figure}

In the complete one-loop energy shift (\ref{OneLoopE}) the constant terms in (\ref{FreqTable}) will drop out and thus do not need to be computed. This is in particular clear, when performing the graded sum  over $\Omega^{ij} (x_n^{ij})$ with the explicit frequencies in (\ref{FreqTable}).

For the general case of not symmetric solutions, we can repeat the above analysis, however the minimal set of required off-shell fluctuation frequencies will generically be larger than two.  

%%%%%%%%%%%%%%%%%%%%%%%%%%%%%%%%%%%%%%%%%
%%%%%%%%%%%%%%%%%%%%%%%%%%%%%%%%%%%%%%%%%

\subsubsection{Exampe: Circular String}

We shall now specialize to the case of  $\mathfrak{su}(2)$ solutions, and then apply these results to the circular string discussed in section \ref{sec:Circular}.  For $\mathfrak{su}(2)$ soltuions, only $\t p_2$ (and $\t p_3$) will be connected by square root cuts (outside the unit circle) and 
\begin{equation}
\t p_2=-\t p_3 \,,\qquad \t p_1=-\t p_4\qquad  \text{and} \qquad  \h p_1=\h p_2 =-\h p_3=-\h p_4 \,,    
\end{equation}
so that we will generically have 6 different frequencies, namely:

\begin{enumerate}
\item{One internal fluctuation corresponding to a pole shared by $\t p_2$ and $\t p_3$ which we denote by
\begin{equation}
\Omega_S(y)= \Omega^{\t2  \t3}(y)
\end{equation}}
\item{Another $S^3$ fluctuation connecting $\t p_1$ and $\t p_4$ 
\begin{equation}
\Omega_{\bar S}(y)= \Omega^{\t1  \t4}(y)
\end{equation}}
\item{Two fluctuations which live in $S^5$ but are orthogonal to the ones in $S^3$,
\begin{equation}
\Omega_{S_{\perp}}(y)= \Omega^{\t1  \t3}(y)= \Omega^{\t1  \t4}(y)
\end{equation} }
\item{Four $AdS_5$ fluctuations
\begin{equation}
\Omega_{A}(y)= \Omega^{\h1  \h3 }(y)= \Omega^{\h1  \h4 }(y)=\Omega^{\h2  \h3 }(y)=\Omega^{\h2  \h4 }(y)
\end{equation} }
\item{Four fermionic excitations which end on either $p_{\t 2}$ or $p_{\t 3}$ (which are the sheets where there are cuts outside the unit circle)
\begin{equation}
\Omega_{F}(y)= \Omega^{\h1  \t3 }(y)= \Omega^{\h2  \t3 }(y)=\Omega^{\t2  \h3 }(y)=\Omega^{\t2  \h4 }(y)
\end{equation} }
\item{Four fermionic poles which end on either $p_{\t 1}$ or $p_{\t 4}$ (which are the sheets where there are cuts inside the unit circle)
\begin{equation}
\Omega_{\bar F}(y)= \Omega^{\h1  \t4 }(y)= \Omega^{\h2  \t4 }(y)=\Omega^{\t1  \h3 }(y)=\Omega^{\t1  \h4 }(y) \,.
\end{equation} }
\end{enumerate}
These expressions apply to any $\mathfrak{su}(2)$ solution, where the cuts are symmetrically arranged (as commented earlier, the more general case follows trivially but may require more "basis fluctuations"). They also apply to higher cut solutions, as exemplified in \cite{Gromov:2008ec}.

We now apply these expressions to the circular string of section \ref{sec:Circular}.
Recall, the quasi-momenta for the circular string in $S^3 \times \mathbb{R}$ depend  on the following parameters of the solution, which are the spin $J$ and winding $m$ repackaged as $\mathcal{J} = J/\sqrt{\lambda}$, $\kappa = \sqrt{\mathcal{J}^2 + m^2}$.
The classical energy is
\begin{equation}\label{Esu2}
\mathcal{E} = {E \over \sqrt{\lambda}} = \sqrt{\mathcal{J}^2 + m^2} \,.
\end{equation}
The classical solution is determined by the quasi-momenta \ref{CircularP}.
The fluctuations were first determined from the sigma-model point of view in  \cite{Frolov:2003tu, Frolov:2004bh}, the exact expansion in terms of $1/\mathcal{J}$ as provided in \cite{SchaferNameki:2006gk} and a derivation of the fluctuation frequencies using the algebraic curve was done in \cite{Gromov:2007aq}. Here we will argue that we only need two frequencies, namely the so-called "internal fluctuations" within the $S^3$ and one $AdS$-fluctuation (which is trivial to obtain).

The off-shell frequencies in the $(\t 2, \t 3)$ and $(\h 2, \h3)$ directions are
\begin{equation}
\begin{aligned}\label{InputOmegas}
\Omega^{\t 2 \t 3} (y) & = {2 m+ n_{\t2\t3} \over  \kappa y}
         			  = {2 m +{p_{\t 2} -p_{\t 3} \over 2 \pi} \over  \kappa y}
			          = \frac{2 \sqrt{m^2 y^2+\mathcal{J}^2}}{\left(y^2-1\right)
                              \sqrt{m^2+\mathcal{J}^2}}			  \cr
\Omega^{\h 2 \h 3} (y) & = {2 \over y^2 -1} \,.
\end{aligned}			
\end{equation}
This will be our only input. 
%{\bf Relate with Vicedo's $q(x)$, seems to be the same as off-shell frequency!}.
We will now demonstrate that the remaining $\mathfrak{su}(2)$ frequencies can be obtained with the methods outlined in the last section. 

The AdS-frequencies are all given by generalizations of (\ref{OmegaMapAdS}) 
\begin{equation}
\begin{aligned}
\Omega^{\h1 \h4} (y) &= -\Omega^{\h2\h3} (1/y) -2
 				  = {2 \over y^2-1} \cr
\Omega^{\h2 \h 4} (y) & = {1\over 2} \left(\Omega^{\h2 \h3} + \Omega^{\h1 \h 4} \right)
				  = {2 \over y^2 -1} \cr
\Omega^{\h1\h3} (y) & = 	-\Omega^{\h2\h4} (1/y) -2
 				  = {2 \over y^2-1}	\,.		
\end{aligned}
\end{equation}
Thus showing the expected agreement of all AdS fluctuation energies.

Let us move to the less trivial $S^5$ fluctuations.
{}From (\ref{FreqTable}) we know
\begin{equation}
\Omega^{\t 1\t4} (y) = -\Omega^{\t 2\t3} (1/y)+\Omega^{\t2\t3}(0)   \,.
\end{equation}
Applied to (\ref{InputOmegas}) we get
\begin{equation}
\Omega^{\t 1\t4} (y)
=
\frac{2 \left(-\mathcal{J} y^2+y \sqrt{m^2+y^2 \mathcal{J}^2}
   +\mathcal{J}\right)}{\left(y^2-1\right) \sqrt{m^2+\mathcal{J}^2}}
= {n_{\t1\t4}  y - 2 \mathcal{J} \over \kappa} \,,
\end{equation}
by recalling that $n_{\t1\t4} = {p_{\t 1}(y)- p_{\t 4} \over 2\pi}$.
The remaining frequencies are obtained by linear combination and inversion
\begin{equation}
\begin{aligned}
\Omega^{\t1\t3} (y) &= {1\over 2} \left(\Omega^{\t 1\t4} + \Omega^{\t2\t3} \right)
                                   =
                                {y ( m+n_{\t1\t3} ) - \mathcal{J} - \sqrt{m^2 y^2 + \mathcal{J}^2}\over \kappa} \cr
\Omega^{\t2\t4} (y) & = -\Omega^{\t1\t3} (1/y) - 2  {\partial \mathcal{E} \over \partial \mathcal{J}}  
={y ( m+n_{\t2\t4} ) - \mathcal{J} - \sqrt{m^2 y^2 + \mathcal{J}^2}\over \kappa}   \,.
\end{aligned}
\end{equation}
Finally we compute the fermion frequencies, which are simply linear combinations
\begin{equation}
\begin{aligned}
\Omega^{\h1 \t 4} (y) &= \Omega^{\t1\t4} (y) + \Omega^{\h1 \t1} (y)
                                = {n_{\h1 \t4} y - \mathcal{J} - \kappa \over \kappa}\cr
\Omega^{\t1 \h3} (y) &= \Omega^{\t1\t4} (y) + \Omega^{\t4 \h3} (y)
             			=  {n_{\h1 \t4} y - \mathcal{J} - \kappa \over \kappa}\,.
\end{aligned}
\end{equation}
Similarly one can check the other fermionic frequencies
\begin{equation}
\Omega_{\h1 \t3} (y) = {1\over 2} \left( \Omega_{\t2\t3} (y) + \Omega_{\h1 \h4}(y)\right)
				= {m+  n_{\h1\t3} \over y \kappa}\,.
\end{equation}
The complete 1-loop energy shift is obtained by
\begin{equation}
\delta E = {1\over 2}\sum_{n \in \mathbb{Z}} \sum_{(ij)} (-1)^{F_{ij}} \Omega^{i j} (x_n^{ij}) \,,
\end{equation}
where $\Omega^{ij}(x_n^{ij})$ are of course now the on-shell frequencies, obtained by evaluating the off-shell frequencies at the position of the poles $x_{n}^{ij}$. Note that the sum can be converted into a contour integral in the $n$-plane (see e.g. \cite{SchaferNameki:2006gk, Gromov:2007aq}), which simplifies the evaluation of the energy shift. 
This is in complete agreement with \cite{Frolov:2003tu, Frolov:2004bh, Gromov:2007aq}.

%%%%%%%%%%%%%%%%%%%%%%%%%%%%%%%
\subsection*{Acknowledgements}

I thank Niklas Beisert for comments on the manuscript and for making this review happening. I would also like to thank the Caltech Theory Group for their generous hospitality. 

%\bibliography{intads}
%\bibliographystyle{nb}
\phantomsection
\addcontentsline{toc}{section}{\refname}
\bibliography{chapters,intads}
\bibliographystyle{nb}

\end{document}